\documentclass[conference]{IEEEtran}

\usepackage{amsmath}
\usepackage{threeparttable}

\usepackage{array}
\usepackage{makecell}

\usepackage{amsmath}
\usepackage{nicefrac}

\usepackage{tikz}
\usepackage{circuitikz}
\usepackage{pgfplots}
\usetikzlibrary{calc}
\usetikzlibrary{shapes}
\usetikzlibrary{matrix}
\usetikzlibrary{arrows}
\usetikzlibrary{decorations.markings}
\usetikzlibrary{spy}
\usetikzlibrary{backgrounds}

\usepackage{pgfplots}
\usepgfplotslibrary{units}

\pgfplotsset{
    legend image with text/.style={
        legend image code/.code={%
            \node[anchor=center] at (0.3cm,0cm) {#1};
        }
    },
}

\IEEEoverridecommandlockouts

\usepackage{todonotes}
\presetkeys{todonotes}{inline}{}
\usepackage{mathrsfs}
\usepackage{amssymb}
\usepackage{gensymb}

\begin{document}

\bstctlcite{IEEEexample:BSTcontrol}

%
\title{Design and Implementation of a Neural Network Aided Self-Interference Cancellation Scheme for Full-Duplex Radios}

%
\author{\IEEEauthorblockN{Yann Kurzo, %
Andreas Burg, %
Alexios Balatsoukas-Stimming %
}%
\IEEEauthorblockA{Telecommunications Circuits Laboratory\\\'Ecole polytechnique f\'ed\'erale de Lausanne, CH-1015 Lausanne, Switzerland}
}

\maketitle

\begin{abstract}
In-band full-duplex systems are able to transmit and receive information simultaneously on the same frequency band. Due to the strong self-interference caused by the transmitter to its own receiver, the use of non-linear digital self-interference cancellation is essential. In this work, we present a hardware architecture for a neural network based non-linear self-interference canceller and we compare it with our own hardware implementation of a conventional polynomial based canceller. We show that, for the same cancellation performance, the neural network canceller has a significantly higher throughput and requires fewer hardware resources.
\end{abstract}


\section{Introduction}
In-band full-duplex (FD) communication has for long been considered to be impractical due to the strong self-interference (SI) caused by the transmitter to its own receiver. However, more recent work on the topic (e.g.,~\cite{Jain2011,Duarte2012,Bharadia2013}) has demonstrated that it is in fact possible to achieve sufficient SI cancellation to make FD systems viable. A portion of the SI is usually first removed in the analog RF domain. As analog cancellation alone is often not sufficient, the residual SI needs to be cancelled in the digital domain. In principle, the residual SI should be easy to cancel since it is produced by a known transmitted signal. In practice, however, the different stages of the transceiver introduce non-linearities to the signal, such as digital-to-analog converter (DAC) and analog-to-digital converter (ADC) non-linearities, IQ imbalance, and power amplifier (PA) non-linearities. Intricate memory polynomial models have to be used in order for the digital SI cancellation to be able to handle the aforementioned non-linearities~(e.g., \cite{Sahai2013,Syrjala2014,Anttila2014,Balatsoukas2015,Korpi2017}). An alternative solution, which uses a neural network (NN) to reconstruct the non-linearities in order to generate the SI cancellation signal, was recently proposed in~~\cite{Balatsoukas2018} and it was shown that it can achieve similar SI cancellation performance with the state-of-the-art polynomial model of~\cite{Korpi2017}, but with much lower computational complexity.

Existing NN hardware accelerators, such as~\cite{Zhang2015,Chen2016}, mainly target applications where both the size of the NN and the number of inputs is very large, and where producing a a few tens of outputs per second is sufficient. Communications applications, on the other hand, use relatively small NNs with few inputs, but need to provide millions of outputs per second. As such, communications applications require vastly different NN hardware accelerator architectures.

\subsubsection*{Contribution}
In this work, we present a hardware implementation of the SI cancellation method proposed in~\cite{Balatsoukas2018} in order to quantify and translate the computational complexity gains over the state-of-the-art polynomial based model of~\cite{Korpi2017} into real-world hardware resource utilization gains. We provide FPGA and ASIC implementation results that clearly demonstrate the significant gains that can be achieved by our proposed NN-based canceller in terms of both the resource utilization and the achieved throughput. To the best of our knowledge, this is the first hardware implementation of a NN-augmented communications system in literature related to the recent resurgence of machine learning for communications.


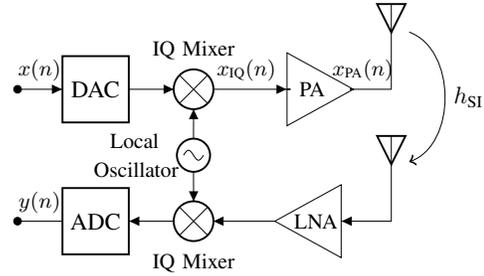
\begin{figure}[t]
  \centering
  \scalebox{0.9}{\begin{circuitikz}[scale=0.65]

	




	\draw (4,0) node[mixer,scale=0.6] (txmixer) {};
	\draw (8.5,0) node[antenna,scale=0.6] (txantenna) {};
	
	\draw (0,0) to[short,*-] ++(0,0) to[twoport,>,t=DAC] (txmixer.west) node[inputarrow]{};
	\draw (txmixer.east) to[short,-] ++(1.75,0) to[amp,>,t=\small{PA}] ++(1.35,0) to[short,-] (txantenna);
	
	\draw (0.5,0) node[above] {\small $x(n)$};
	\draw (txmixer)+(0,0.5) node[above] {\small{IQ Mixer}};
	\draw (txmixer)+(1.2,0) node[above] {\small $x_{\text{IQ}}(n)$};
	\draw (txantenna)+(-0.65,0) node[above] {\small $x_{\text{PA}}(n)$};
	
	\draw (8.5,-3) node[antenna,scale=0.6] (rxantenna) {};
	\draw (4,-3) node[mixer,scale=0.6] (rxmixer) {};
	
	\draw (rxantenna) to[short,-] ++(-1.15,0) to[amp,>,t={\rotatebox[origin=c]{180}{\small{LNA}}}] ++(-1.5,0) to[short,-] (rxmixer.east) node[inputarrow,rotate=180]{};
	\draw (rxmixer.west) to[twoport,>,t=ADC] (0,-3) to[short,-*] (0,-3);
	\draw [->] (txantenna)+(0.4,1.3) to[thick, out=-20, in=20, edge node={node [right] {$h_{\text{SI}}$}}]  ($(rxantenna) + (0.4,1.35)$);
	
	\draw (0.5,-3) node[above] {\small $y(n)$};
	\draw (rxmixer)+(0,-0.55) node[below] {\small{IQ Mixer}};
	
	\draw (4.375,-1.5) node[oscillator,scale=0.5] (ref) {};
	\draw (ref.south) to[short,-] (rxmixer.north) node[inputarrow,rotate=270]{};
	\draw (ref.north) to[short,-] (txmixer.south) node[inputarrow,rotate=90]{};
	\draw (ref)+(-0.6,0) node[left,text width=1.2cm,align=center] {\small{Local\\Oscillator}};
	
\end{circuitikz}}
  \caption{Simplified wireless transceiver block diagram.}\label{fig:block}
  \vspace{-0.1cm}
\end{figure}

\section{Digital Self-Interference Cancellation}\label{sec:background}
A basic block diagram of a full-duplex wireless transceiver is shown in Fig.~\ref{fig:block}. If we assume, for simplicity, that there is no signal-of-interest from a remote node and no thermal noise, then the received signal $y(n)$ is the SI signal. The goal of digital SI cancellation is to reproduce an accurate copy of $y(n)$, denoted by $\hat{y}(n)$, based on the transmitted baseband signal $x(n)$. This signal is then subtracted from $y(n)$, so that the residual SI signal is $y_c(n) = y(n) - \hat{y}(n)$. If $\hat{y}(n)$ is reconstructed perfectly, then the SI can be cancelled entirely and $y_c(n) = 0$. In practice, however, due to the presence of thermal noise and transceiver non-linearities, perfect SI cancellation is difficult to achieve.

\subsubsection{Polynomial Non-Linear Cancellation}
A state-of-the-art polynomial SI cancellation model, which can effectively suppress IQ imbalance and PA non-linearities, was described in~\cite{Korpi2017}. Specifically, it was shown that an accurate SI cancellation signal $\hat{y}(n)$ can be obtained as:
\begin{align}
  \hat{y}(n) & = \sum _{\substack{p=1,\\p \text{ odd}}}^P \sum_{q=0}^p\sum_{l=0}^{L-1}\hat{h}_{p,q}(l) \underbrace{x(n-l)^{q}x^*(n-l)^{p-q}}_{\text{basis functions}}, \label{eq:poly}
\end{align}
where $x(n)$ is the transmitted digital baseband signal, $L$ corresponds to the overall memory of the system, $P$ is the non-linearity order, and $\hat{h}_{p,q}$ are estimated parameters that can be obtained using, e.g., least-squares estimation.
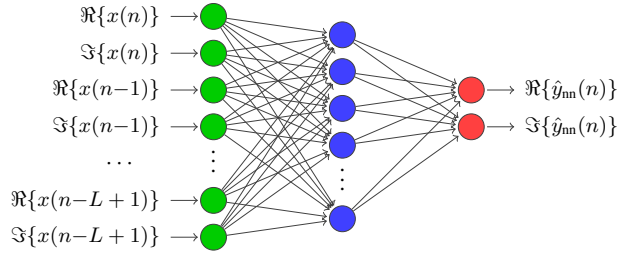
\begin{figure}[t]
  \centering
  \scalebox{0.89}{\def\layersep{3.5cm}
\def\twolayersep{7cm}
\def\inputnodes{6}
\def\hiddennodes{5}
\def\outputnodes{2}

\begin{tikzpicture}[shorten >=1pt,->,draw=black!75, node distance=\layersep, scale=0.55]
    \tikzstyle{every pin edge}=[<-,shorten <=1pt]
    \tikzstyle{neuron}=[circle,draw=black!75,fill=black!75,minimum size=11pt,inner sep=0pt]
    \tikzstyle{input neuron}=[neuron, fill=black!20!green];
    \tikzstyle{output neuron}=[neuron, fill=red!75];
    \tikzstyle{hidden neuron}=[neuron, fill=blue!75];
    \tikzstyle{annot} = [text width=10em, text centered]

	\node[input neuron, pin=left: {\small$\Re{\left\{x(n)\right\}}$}] (I-1) at (0,-1) {};
	\node[input neuron, pin=left: {\small$\Im{\left\{x(n)\right\}}$}] (I-2) at (0,-2) {};
	\node[input neuron, pin=left: {\small$\Re{\left\{x(n{-}1)\right\}}$}] (I-3) at (0,-3) {};
	\node[input neuron, pin=left: {\small$\Im{\left\{x(n{-}1)\right\}}$}] (I-4) at (0,-4) {};
	\node at (-2.5,-5) {$\cdots$};
	\node at (0,-4.75) {$\vdots$};
	\node[input neuron, pin=left: {\small$\Re{\left\{x(n{-}L+1)\right\}}$}] (I-5) at (0,-6) {};
	\node[input neuron, pin=left: {\small$\Im{\left\{x(n{-}L+1)\right\}}$}] (I-6) at (0,-7) {};

	\pgfmathsetmacro{\limit}{\hiddennodes-1}
	\foreach \name / \y in {1,...,\limit}
		\path[yshift=-0.5cm]
			node[hidden neuron] (H-\name) at (\layersep,-\y cm) {};
	\node at (\layersep,-5.25) {$\vdots$};
	\path[yshift=-0.5cm]
		node[hidden neuron] (H-\hiddennodes) at (\layersep,-6 cm) {};

	\foreach \name / \y in {1,...,\outputnodes}
		\path[yshift=-2cm]
			node[output neuron,pin={[pin edge={->}]right:\ifodd\y {\small $\Re{\left\{\hat{y}_{\text{nn}}(n)\right\}}$} \else {\small $\Im{\left\{\hat{y}_{\text{nn}}(n)\right\}}$} \fi}] (O-\name) at (\twolayersep,-\y) {};

    \foreach \source in {1,...,\inputnodes}
        \foreach \dest in {1,...,\hiddennodes}
			\path (I-\source) edge (H-\dest);

    \foreach \source in {1,...,\hiddennodes}
		\foreach \dest in {1,...,\outputnodes}
			\path (H-\source) edge (O-\dest);

\end{tikzpicture}
}
  \caption{Example of a neural network for the reconstruction of the non-linear component of the SI signal~\cite{Balatsoukas2018}.}\label{fig:nn}
  \vspace{-0.1cm}
\end{figure}

\subsubsection{Neural Network Non-Linear Cancellation}
\begin{figure}[t]
  \centering
  \includegraphics[width=1.0\columnwidth]{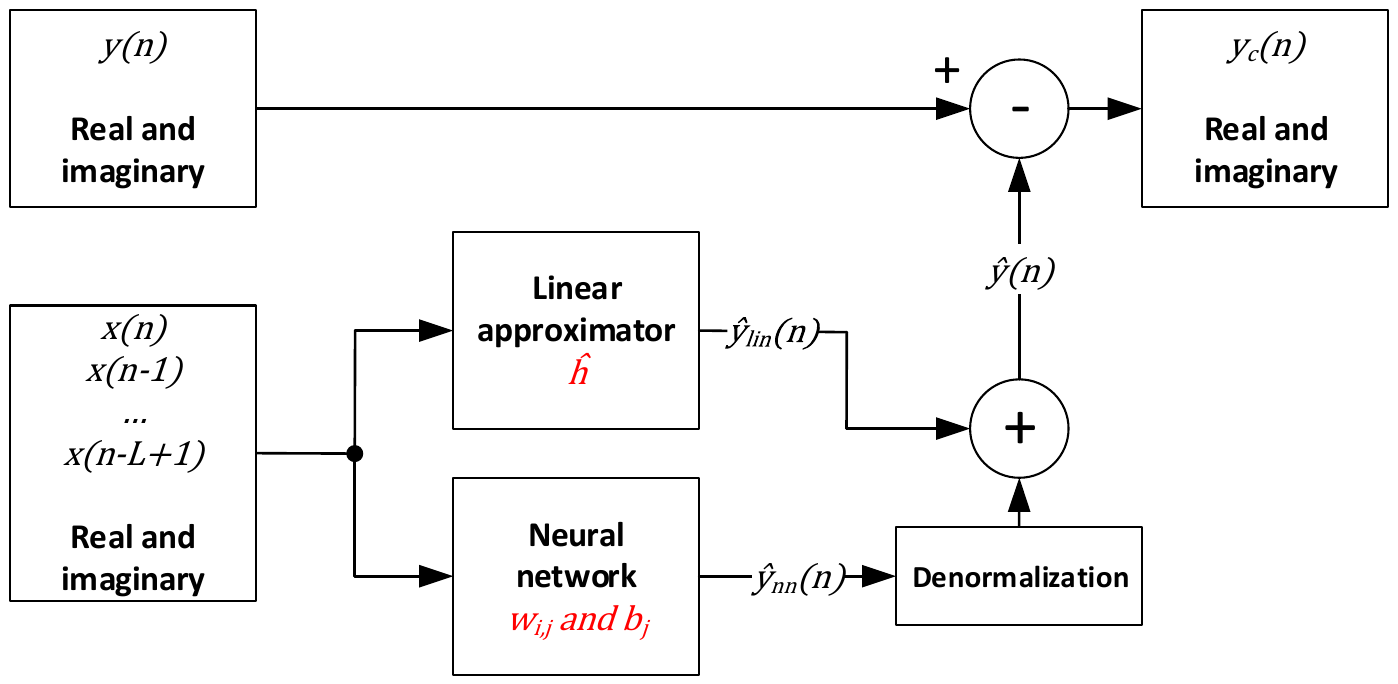}
  \caption{Neural network based SI cancellation.}\label{fig:si-canceller-nn}
  \vspace{-0.1cm}
\end{figure}

The NN-based method of~\cite{Balatsoukas2018} uses two steps, as illustrated in Fig.~\ref{fig:si-canceller-nn}. First, standard linear cancellation is used in order to reconstruct the linear component of the SI, denoted by $\hat{y}_{\text{lin}}(n)$:
\begin{align}
  \hat{y}_{\text{lin}}(n) & = \sum_{l=0}^{L-1}\hat{h}(l) x(n-l), \label{eq:linear}
\end{align}
where $\hat{h}$ are estimated parameters that can be obtained using, e.g., least-squares estimation. A two-layer real-valued neural network, shown in Fig.~\ref{fig:nn}, generates the non-linear part of the SI cancellation signal, denoted by $\hat{y}_{\text{nn}}(n)$. Finally, the two components are added in order to create the SI cancellation signal $\hat{y}(n) = \hat{y}_{\text{lin}}(n) + \hat{y}_{\text{nn}}(n)$. The denormalization step in Fig.~\ref{fig:si-canceller-nn} is necessary because the NN learns to reproduce a normalized (i.e., zero-mean and unit-variance) version of $\hat{y}_{\text{nn}}$, as this generally improves the convergence of NN training.

\subsubsection{Computational Complexity}
Assuming that each complex multiplication can be implemented using three real multiplications and five real additions and that each complex addition can be implemented using two real additions, the total number of real multiplications and additions that are required by the polynomial canceller is~\cite{Balatsoukas2018}\footnote{We note that the expression for $N_{\text{ADD,poly}}$ in our previous work of~\cite{Balatsoukas2018} erroneously ignored the five real additions that are required to implement each complex multiplication. As such, the actual complexity of the polynomial canceller is even higher than that reported in~\cite{Balatsoukas2018}.}:
\begin{align}
	N_{\text{ADD,poly}}	& = \frac{7}{4}L\left(P+1\right)\left(P+3\right)-2,\\
	N_{\text{MUL,poly}}	& = \frac{3}{4}L\left(P+1\right)\left(P+3\right).
\end{align}
The number of real multiplications and additions that are required by the NN canceller is~\cite{Balatsoukas2018}:
\begin{align}
	N_{\text{ADD,NN}}	& = (2L+3)N_h + (7L -2),\\
	N_{\text{MUL,NN}}	& = (2L+2)N_h + 3L,
\end{align}
where the second term in both expressions comes from the linear canceller. The complexity expressions for the two methods can not be compared directly because they contain different sets of parameters. In order to perform a fair comparison we select values for $L$, $P$, and $N_h$ so that the two methods have the same SI cancellation performance in Section~\ref{sec:results}.

\section{Hardware Architecture}\label{sec:hardware}

\begin{figure}[t]
  \centering
  \includegraphics[width=1.0\columnwidth]{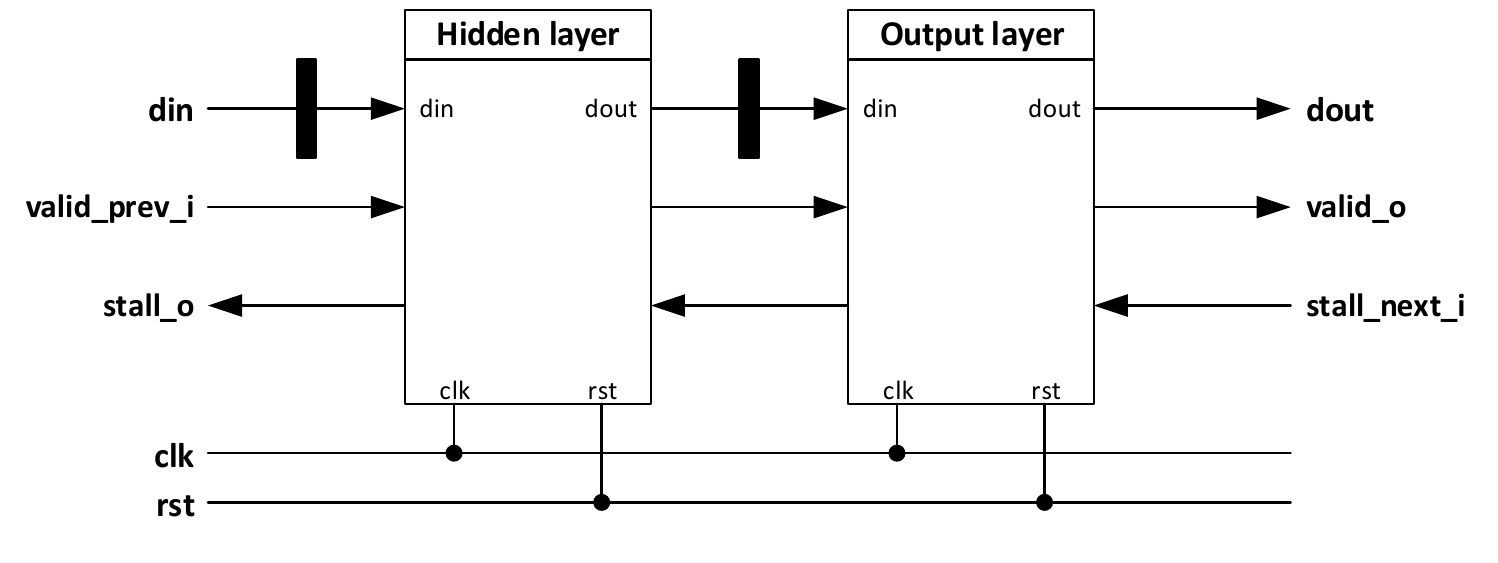}
  \caption{Macro-pipeline architecture of the two-layer neural network.}\label{fig:pipeline-archi}
  \vspace{-0.1cm}
\end{figure}

In this section, we describe a hardware architecture that implements the NN-based SI canceller of~\cite{Balatsoukas2018}. We first give a global overview of the architecture, which is followed by a more detailed explanation of each component. As shown in Fig.~\ref{fig:pipeline-archi}, we map each layer of the NN to a macro-pipeline stage that requires several clock cycles to compute its outputs. Each macro-pipeline stage can start its computations as soon as valid outputs from the previous pipeline stage become available.

\subsection{Macro-Pipeline Architecture}
\begin{figure}[t]
  \centering
  \includegraphics[width=1\columnwidth]{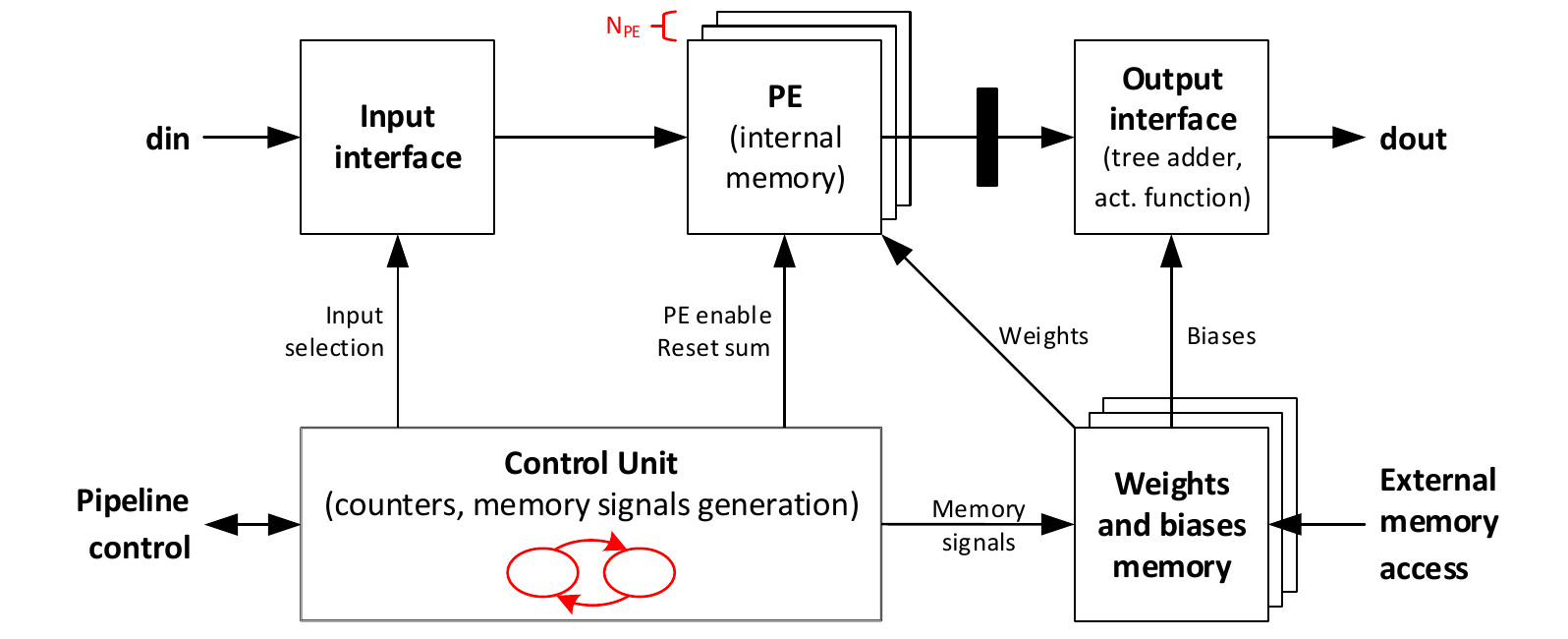}
  \caption{Block diagram of the macro-pipeline stage architecture.}\label{fig:dense-layer-archi}
  \vspace{-0.1cm}
\end{figure}

Let $N_I$ and $N_n$ denote the number of inputs per neuron (which is equal to the number of neurons of the previous layer) and the number of neurons for a given NN layer, respectively. The goal of a macro-pipeline stage is to process each neuron of its corresponding layer by computing the following outputs:
\begin{align}
  o_j = f \left( b_j + \sum_{i=0}^{{N_I-1}} w_{i,j} x_{i} \right), \quad j \in \{0,\hdots,{N_n-1}\}, \label{eq:dense-layer}
\end{align}
where $x_i$ are the inputs, $w_{i,j}$ are the \emph{weights}, $b_j$ are the \emph{biases}, and $f(x)$ is a non-linear activation function (\cite{Balatsoukas2018} uses a ReLU activation function). The architecture of each macro-pipeline stage is shown in more detail in Fig.~\ref{fig:dense-layer-archi}. More specifically, each macro-pipeline stage contains an input interface, an array of $N_{\text{PE}}$ processing elements (PEs), a weights-and-biases memory, a control unit, and an output interface. We note that all weights, biases, and partial sums have a common bit-width of $Q$ bits and saturation is used in case of an overflow.

The $N_{\text{PE}}$ PEs, whose internal structure is shown in Fig.~\ref{fig:rpe-archi}, can be used to compute~\eqref{eq:dense-layer} over multiple clock cycles using one of two possible schedules. In the neuron-by-neuron (NBN) schedule, neurons are processed sequentially and each of the $N_{\text{PE}}$ PEs computes a part of the sum in~\eqref{eq:dense-layer} for a given neuron $j$. In the input-by-input (IBI) schedule, on the other hand, the layer inputs $x_{i}$ are processed sequentially and the $N_{\text{PE}}$ PEs update the sum in~\eqref{eq:dense-layer} with the term $w_{i,j} x_{i}$ for $N_{\text{PE}}$ neurons in parallel. When an NBN macro-pipeline stage is followed by an IBI macro-pipeline stage, the IBI stage can already start performing computations once the output of the first neuron of the NBN stage has been computed, thus masking a significant part of the latency and reducing the number of interconnects between the two stages. Since the exact architecture of each macro-pipeline stage depends on the processing schedule, we describe the details of the corresponding architectures separately in the next two sections.

\subsection{Neuron-by-Neuron Macro-Pipeline Architecture}

\subsubsection{Input Interface}
The input interface consists of $N_{\text{PE}}$ multiplexers, which route each input to the correct PE.

\subsubsection{Processing Elements}
In the NBN schedule, each PE is only associated with a single neuron, meaning that only a single partial sum needs to be stored. Thus, the PEs are simple multiply-and-accumulate (MAC) units and the memory shown in Fig.~\ref{fig:rpe-archi} is in fact a single $Q$-bit register.

\subsubsection{Control Unit}
The main tasks of the control unit are to distribute the computations to the PEs and to stall the computations when no valid inputs are available or when the following macro-pipeline stage is not ready to accept new outputs. The computations are dispatched to the PEs as follows. When $N_{\text{PE}} \leq N_I$, all $N_{\text{PE}}$ PEs are used to process a single neuron at a time and $N_n\left\lceil\frac{N_I}{N_{\text{PE}}}\right\rceil$ clock cycles are required to process all neurons. When $N_{\text{PE}} > N_I$, we constrain $N_{\text{PE}}$ so that $N_{\text{PE}} = kN_I,~k \in \mathbb{N},$ meaning that $k$ neurons are processed in parallel and $\left\lceil\frac{N_nN_I}{N_{\text{PE}}}\right\rceil$ clock cycles are required to process all neurons.

\subsubsection{Weights and Biases Memories}
The weights and biases memories are used to store $w_{i,j}$ and $b_j$ and they can be written externally to re-configure the NN. The weights are organized in a memory that is $N_{\text{PE}}Q$ bits wide so that all PEs can be provided with data in parallel. A single word of the weights memory contains $N_{\text{PE}}$ weight values corresponding to $k$ different neurons. The biases memory, on the other hand, has a bit-width of $kQ$ bits.

\subsubsection{Output Interface}
The output interface adds the partial sums from the $N_{\text{PE}}$ PEs using an adder tree, it adds the biases, and it applies the non-linear activation function for each of the $k$ neurons that are being processed in parallel. A register is added between the PEs and the output interface in order to reduce the critical path of the architecture. Moreover, the output interface forwards the outputs of the $k$ neurons that are processed in parallel to the next macro-pipeline stage.


\subsubsection{Latency}
If $N_{\text{PE}}$ is chosen carefully so that $\frac{N_I}{N_{\text{PE}}}$ and $\frac{N_nN_I}{N_{\text{PE}}}$ are always integers, then it takes
\begin{align}
 \mathcal{L} &  = \frac{N_nN_I}{N_{\text{PE}}} + 1
\end{align}
clock cycles to produce all outputs of a NN layer. However, one full set of outputs for a NN layer is actually produced every $\frac{N_nN_I}{N_{\text{PE}}}$ cycles, so that the throughput of the NBN macro-pipeline stage is
\begin{align}
 \mathcal{T}&  = \frac{N_{\text{PE}}}{N_nN_I}
\end{align}
Moreover, the first $k$ outputs of an NBN macro-pipeline stage become available after
\begin{align}
 \mathcal{L}_f & = \frac{N_I}{N_{\text{PE}}} + 1
\end{align}
clock cycles and after that $k$ new outputs are produced  in every clock cycle. This means that a potential IBI macro-pipeline stage that follows can already start its computations after $\mathcal{L}_f$ clock cycles and that only $k \leq N_{n}$ outputs need to be forwarded to the next stage in each clock cycle.

\subsection{Input-by-Input Macro-Pipeline Architecture}
\subsubsection{Input \& Output Interface}
The input and output interfaces are similar to that of the NBN macro-pipeline stage, the main difference being that the IBI output interface forwards the outputs of all $N_n$ neurons that are processed in parallel to the next macro-pipeline stage.

\subsubsection{Processing Elements}
\begin{figure}[t]
  \centering
  \includegraphics[width=1.0\columnwidth]{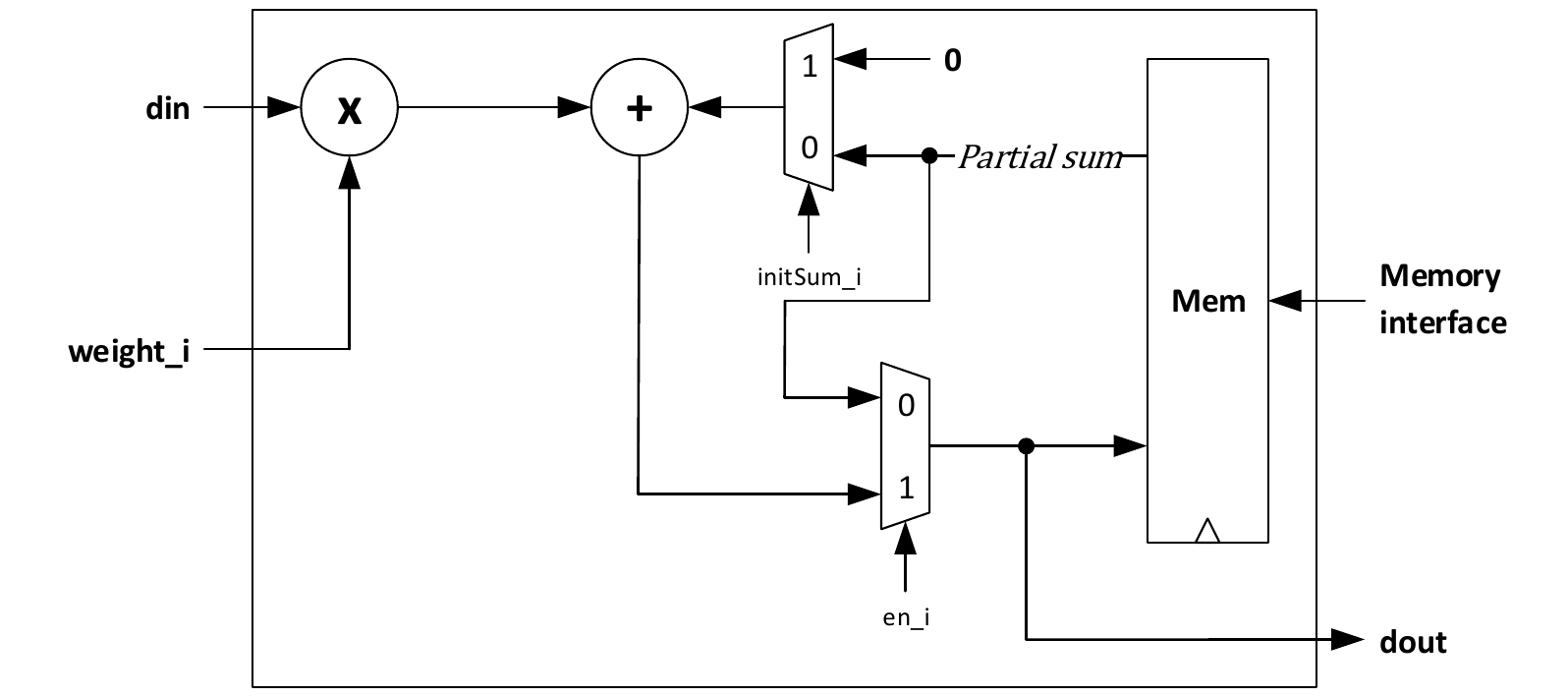}
  \caption{Detailed view of the PE architecture that is used by both the NBN and the IBI macro-pipeline stages.}\label{fig:rpe-archi}  
  \vspace{-0.1cm}
\end{figure}

In the IBI schedule, each PE can be associated with multiple neurons, meaning that several partial sums potentially need to be stored. Thus, the PEs are MAC units and the memory shown in Fig.~\ref{fig:rpe-archi} has a dimension of $\left\lceil\frac{N_n}{N_{\text{PE}}}\right\rceil \times Q$ bits.

\subsubsection{Control Unit}
In the IBI schedule, when $N_{\text{PE}} \leq N_n$, all $N_{\text{PE}}$ PEs are used to update the $N_n$ neurons sequentially with the new input value $x_i$ and $N_I\left\lceil\frac{N_n}{N_{\text{PE}}}\right\rceil$ clock cycles are required to process all neurons. When $N_{\text{PE}} > N_n$, we constrain $N_{\text{PE}}$ so that $N_{\text{PE}} = kN_n,~k \in \mathbb{N},$ meaning that $k$ inputs are processed in parallel and $\left\lceil\frac{N_nN_I}{N_{\text{PE}}}\right\rceil$ clock cycles are required to process all neurons.

\subsubsection{Weights and Biases Memories}
The weights and biases memories are similar to those of the NBN macro-pipeline stage. A single word of the weights memory contains $N_{\text{PE}}$ weight values corresponding to $k$ different neurons. The biases memory, on the other hand, has a bit-width of $N_nQ$ bits.

\subsubsection{Latency}
Similarly to the NBN schedule, if $N_{\text{PE}}$ is chosen carefully so that $\left\lceil\frac{N_n}{N_{\text{PE}}}\right\rceil$ and $\left\lceil\frac{N_nN_I}{N_{\text{PE}}}\right\rceil$ are always integers, then the latency and the throughput are
\begin{align}
 \mathcal{L}&  = \frac{N_nN_I}{N_{\text{PE}}} + 1, \quad \text{ and } \quad \mathcal{T} = \frac{N_{\text{PE}}}{N_nN_I},
\end{align}
respectively. Moreover, all $N_n$ outputs of an IBI macro-pipeline stage become available simultaneously after
\begin{align}
 \mathcal{L}_f & =  \frac{N_nN_I}{N_{\text{PE}}} + 1 \quad \text{clock cycles.}
\end{align}

\subsection{Overall Neural Network Canceller Architecture}
The overall architecture for the two-layer NN of~\cite{Balatsoukas2018} consists of two macro-pipeline stages, one for the hidden layer and one for the output layer, and pipeline registers are added between the macro-pipeline stages. The hidden layer uses an NBN macro-pipeline stage, while the output layer uses an IBI macro-pipeline stage. For the hidden layer, we have $N_I = 2L$ and $N_n = N_h$, while for the output layer we have $N_I = N_h$ and $N_n = 2$. The $N_I = 2L$ inputs of the first macro-pipeline stage that implements the computations of the hidden layer are assumed to all be available in parallel. The number of PEs instantiated for the hidden layer and the output layer is $N_{\text{PE},h}$ and $N_{\text{PE},o}$, respectively. The computations for the linear canceller are done in parallel with the NN by instantiating a standard complex FIR filter. If we denote the throughput of the hidden and the output macro-pipeline stages by $\mathcal{T}_h$ and $\mathcal{T}_o$, respectively, then the throughput of the two-layer NN architecture is
\begin{align}
  \mathcal{T} & = \min\left(\mathcal{T}_h,\mathcal{T}_o\right).
\end{align}
Finally, we note that we constrain the denormalization step shown in Fig.~\ref{fig:si-canceller-nn} to scaling with powers of two, which can be implemented efficiently with simple shifting operations, both during training and during inference.

\section{FPGA and ASIC Implementation Results} \label{sec:results}

In this section, we present implementation results for the NN-based canceller and we compare it with a polynomial canceller. Since, to the best of our knowledge, there are no published implementations of polynomial cancellers in the literature, we provide our own reference implementation. Due to space limitations, we do not describe the implementation in detail, but it is largely based on the NN architecture since the main computational task of the polynomial canceller is similar to the NN canceller, i.e., to compute a weighted sum. The main differences are that the input interface also computes the basis functions and that the PEs operate directly on complex numbers. Each complex PE of the polynomial canceller is implemented using three real multipliers.

\subsection{Comparison Setup}
In order to provide a fair comparison between the NN-based SI canceller and the polynomial canceller, we select $L$, $N_h$, $P$, and the quantization bit-width $Q$ so that the fixed-point performance of the two cancellers is as similar as possible. For performance evaluation, we used the same dataset that was used in~\cite{Balatsoukas2018}, which consists of a $10$~MHz QPSK-modulated OFDM signal sampled at $20$~MHz that is generated using the testbed described in~\cite{Balatsoukas2013a} and \cite{Balatsoukas2013b}.

\begin{table}[t]
  \centering
  \caption{Comparison of NN-based and polynomial cancellers.}\label{tab:perf}
  \begin{tabular}{l|cc}
                          & Polynomial & Neural Network \\
    \hline
    Cancellation (dB)     & $-44.8$    & $-44.4$ \\
    Real Parameters       & $520$      & $550$ \\
    Real Multiplications  & $780$      & $543$ \\
    Real Additions        & $1818$     & $611$ \\
  \end{tabular}
  \vspace{-0.1cm}
\end{table}

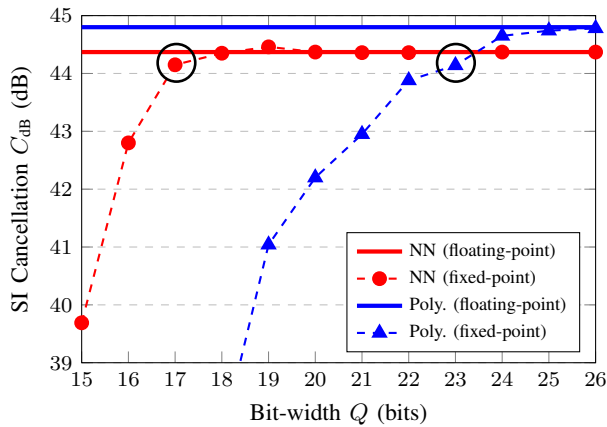
\begin{figure}[t]
  \centering
  \begin{tikzpicture}

	\pgfplotsset{grid style={dashed}}

	\begin{axis}[
		width = 0.95\columnwidth,
		height = 0.7\columnwidth,
		xlabel = {Bit-width $Q$ (bits)},
		ylabel = {SI Cancellation $C_{\text{dB}}$ (dB)},
		ylabel near ticks,
		xlabel near ticks,
		xtick distance=1,
		ytick distance=1,
		label style={font=\small},
		tick label style={font=\footnotesize},
		xmin = 15, xmax = 26,
		ymin = 39, ymax = 45,
		ymajorgrids,
		legend pos = south east,
		legend style={font=\scriptsize,cells={align=left}},
		legend columns=1,
		legend cell align=left,
		legend entries={NN (floating-point),
                    NN (fixed-point),
                    Poly. (floating-point),
                    Poly. (fixed-point)}
	]

		\addplot[red, ultra thick, solid] table[x index = 2, y index = 3] {fig/data/bitwidth_power_nn.dat};
		\addplot[red, thick, dashed, mark=*, mark options={solid, ultra thick}] table[x index = 0, y index = 1] {fig/data/bitwidth_power_nn.dat};

		\addplot[blue, ultra thick, solid] table[x index = 2, y index = 3] {fig/data/bitwidth_power_poly.dat};
		\addplot[blue, thick, dashed, mark=triangle*, mark options={solid, ultra thick}] table[x index = 0, y index = 1] {fig/data/bitwidth_power_poly.dat};

	\end{axis}

	\draw[line width=1] (1.26,3.99) circle (0.25);
	\draw[line width=1] (4.975,3.99) circle (0.25);

\end{tikzpicture}%
  \caption{SI cancellation as a function of the datapath bit-width $Q$.}\label{fig:quantization}
  \vspace{-0.1cm}
\end{figure}

For $L=13$, $P=7$, and $N_h=18$, the performance of the two cancellers is very similar, as can be seen in Table~\ref{tab:perf}. In Fig.~\ref{fig:quantization}, we show the cancellation performance for the NN-based canceller and the polynomial based canceller as a function of $Q$. We observe that, for the same cancellation performance, the NN-based canceller generally requires a lower quantization bit-width $Q$. For the hardware implementation results, we choose $Q=17$ for the NN-based canceller and $Q=23$ for the polynomial canceller so that the two cancellers have the same fixed-point cancellation performance.

We set $N_{\text{PE},h} = 52$ and $N_{\text{PE},o} = 4$ for the NN-based canceller so that $\mathcal{T}_h = \mathcal{T}_o = \nicefrac{1}{9}$, meaning that the macro-pipeline is perfectly balanced and one cancellation sample is output every $9$ clock cycles. Furthermore, $2$ complex PEs are instantiated for the NN-based canceller in order to perform the linear cancellation step in the same time. For the polynomial canceller, we use $N_{\text{PE},h} = 20$ complex PEs so that the $260$ complex multiplications required to compute $\eqref{eq:poly}$ for $L=13$ and $P=7$ can be carried out in $13$ clock cycles, which means that one cancellation sample is output every $13$ clock cycles.

\subsection{Implementation Results}
The placed-and-routed implementation results on a Xilinx Virtex-7 FPGA are given in Table~\ref{tab:fpga-res}. We observe that the NN-based canceller has significantly lower resource utilization than the polynomial canceller and a $96$\% higher throughput. The higher throughput of the NN-based canceller comes both from a lower number of cycles per sample and from a higher operating frequency compared to the polynomial canceller. We also note that the polynomial canceller requires approximately two times more DSP slices than the NN-based canceller. This happens because the DSP slices on Xilinx Virtex-7 FPGAs do not support multiplications between two $Q=23$-bit values and two DSP slices have to be instantiated for each multiplication in the polynomial canceller.

\begin{table}[t]
  \centering
  \caption{FPGA Implementation Results (Virtex-7 XC7VX485TFFG1161).}\label{tab:fpga-res}
  \begin{tabular}{l|cc}
                      & Polynomial   & Neural Network \\
    \hline
    LUT (logic)       & $6710/303.6$k ($2.21$\%) & $2831/303.6$k ($0.93$\%) \\
    LUT (RAM)         & $1638/130.8$k ($1.25$\%) & $1678/130.8$k ($1.28$\%) \\
    Registers         & $3922/607.2$k ($0.65$\%) & $2625/607.2$k ($0.43$\%) \\
    DSP Slices        & $132/2.8$k    ($4.71$\%) & $62/2.8$k ($2.21$\%) \\
    Frequency (MHz)   & $67.2$  & $92.0$ \\
    Throughput (MS/s) & $5.2$   & $10.2$
  \end{tabular}
  \vspace{-0.1cm}
\end{table}

The fully placed-and-routed ASIC implementation results using a 28~nm FD-SOI technology are shown in Table~\ref{tab:asic}. We observe that the NN-based canceller has a $60$\% better throughput and that it occupies $11$\% less area than the polynomial canceller, leading to an $81$\% better hardware efficiency. Similarly to the FPGA results, the better throughput of the NN-based canceller comes both from a lower number of cycles per sample and from a higher operating frequency compared to the polynomial canceller.

\begin{table}[t]
  \centering
  \caption{ASIC Implementation Results (28~nm FD-SOI).}\label{tab:asic}
  \begin{threeparttable}
  \begin{tabular}{l|cc}
                              & Polynomial  & Neural Network \\
    \hline
    Area (mm$^2$)             & $0.36$      & $0.32$ \\
    Frequency (MHz)           & $226$       & $250$ \\
    Throughput (MS/s)         & $17.4$      & $27.8$ \\
    Efficiency (MS/s/mm$^2$)  & $48$        & $87$ \\
  \end{tabular}
  \end{threeparttable}
  \begin{tablenotes}
    \item PAR results using slow corners, $0.7$~V voltage, $125$\degree~C temperature.
  \end{tablenotes}
  \vspace{-0.1cm}
\end{table}

\section{Conclusion}
In the paper, we described a high-throughput hardware architecture for a NN-based self-interference cancellation scheme for full-duplex radios. Our implementation results show that the NN-based canceller has a lower computational complexity and that a $22$\% lower datapath quantization bit-width to achieve the same cancellation performance as a polynomial cancellation scheme. The NN-based canceller thus requires significantly fewer resources on an FPGA and achieves an $81$\% better hardware efficiency than the polynomial canceller when implemented for an ASIC target.

\section{Acknowledgment}
The authors gratefully acknowledge the support of NVIDIA Corporation with the donation of the Titan Xp GPU used for this research. This work has been supported by the Swiss National Science Foundation under grant \#182621.

\bstctlcite{IEEEexample:BSTcontrol}
\bibliographystyle{IEEEtran}
\bibliography{IEEEabrv,bibliography}

\end{document}